\documentclass[12pt,preprint]{aastex} 
 
\tighten 
 
\def\lsim{\lower.5ex\hbox{$\; \buildrel < \over \sim \;$}} 
\def\gsim{\lower.5ex\hbox{$\; \buildrel > \over \sim \;$}} 
\def\lax {\ifmmode{_<\atop^{\sim}}\else{${_<\atop^{\sim}}$}\fi} 
\def\gax {\ifmmode{_>\atop^{\sim}}\else{${_>\atop^{\sim}}$}\fi} 
 
\def\gtorder{\mathrel{\raise.3ex\hbox{$>$}\mkern-14mu 
\lower0.6ex\hbox{$\sim$}}} 
\def\ltorder{\mathrel{\raise.3ex\hbox{$<$}\mkern-14mu 
\lower0.6ex\hbox{$\sim$}}}

\def\pmb#1{\setbox0=\hbox{#1}%
\kern-0.015em\copy0\kern-\wd0 
\kern0.03em\copy0\kern-\wd0 
\kern-0.015em\raise0.0433em\box0 } 
 
\begin{document} 
 
\title{Timing  and Spectral Properties of X-ray Emission from the Converging 
Flows onto Black hole: Monte-Carlo Simulations} 

\author{Philippe Laurent\altaffilmark{1},
Lev Titarchuk \altaffilmark{1,2,3}} 
\altaffiltext{1}{CEA, DSM/DAPNIA/SAp, Centre d'Etudes de Saclay, 91191
Gif-sur-Yvette Cedex, France; plaurent@cea.fr}

\altaffiltext{2}{George Mason University/Center for Earth Observing and
Space Research, Fairfax VA 22030-4444;}

\altaffiltext{3}{US Naval Research Laboratory,  Space Science
Division, 4555 Overlook Avenue, SW, Washington,
DC 20375-5352; lev@xip.nrl.navy.mil}

\vskip 0.5 truecm 
 

\begin{abstract} 
We demonstrate that a X-ray spectrum of a converging inflow (CI) onto a black hole is the sum of 
a thermal (disk) component and the convolution of some fraction of this component with  
the Comptonization spread (Green's) function. The latter  component is seen as an 
extended power law at energies much higher than the characteristic energy of the soft photons. 
We show that the high energy photon production (source function) in the CI atmosphere 
is distributed  with the characteristic maximum at about the photon bending radius, $1.5r_{\rm S}$, 
independently of the seed 
(soft) photon distribution. We show that high frequency oscillations 
of the soft photon source in  this region lead to  the oscillations 
of the high energy part 
of the spectrum but not {of} the thermal component. 
The high frequency oscillations of the inner region are not significant 
in the thermal component of the  spectrum.   
We further demonstrate that Doppler and recoil effects 
(which are responsible for the formation of the CI spectrum) 
are related to the hard (positive) and soft (negative) time lags between the soft and hard 
photon energy channels  respectively. 
\end{abstract} 
 
 
 
\section{Introduction} 
 
Accreting stellar-mass black holes (BH) in 
Galactic binaries exhibit 
so called high-soft and low-hard spectral states (e.g. Borozdin et al. 1999, hereafter BOR99). 
An increase in the soft blackbody luminosity component leads to the 
appearance of an extended power law. An important observational fact is 
that this effect is seen as a persistent phenomenon only in BH candidates, 
and thus it is apparently a {\it unique} black hole signature. 
Although in Neutron star (NS) systems  similar power law components are 
detected in the intermediate stages (Strickman \& Barret 1999; Iaria et al. 2000; 
Di Salvo et al. 2001), they are of a  transient nature, disappearing with 
increasing luminosity  (Di Salvo et al.). 
 
It thus seems a reasonable assumption that the unique spectral signature 
of the soft state of BH binaries is directly tied to the black hole event 
horizon. This is the primary motivation for the Bulk Motion Comptonization Model 
(BMC) introduced in several previous papers, and recently applied with striking 
success to a substantial body of observational data (Shrader \& Titarchuk 1998; 
BOR99; Titarchuk \& Shrader 2002, hereafter TS02). A complete theory of 
BH accretion must, however, be also able to accommodate in a 
natural manner a growing number of observational traits exhibited in the temporal 
domain. For example, it is now well established that BH X-ray binaries 
exhibit quasi-periodic oscillation (QPO) phenomena in three  
frequency domains: $\sim 0.1$ Hz, ($\sim 1-10$ Hz) and 
($\sim 10^2$ Hz). 
The $\sim 10^2-$Hz QPOs seem to occur 
during periods of flaring, and when the spectra (although in the high-soft state) 
tend to be relatively hard, i.e. the relative importance of the power-law component 
 with respect to the thermal.  Furthermore, the QPO 
amplitudes increase with energy, that is there is a higher degree of 
modulation of the signal in the hard-power law than in the thermal excess component. 
In addition to the QPO phenomena, it has been noted by Nowak et al. 
(2000) (also see Cui et al. 2000) that measures of the coherence 
between the intensity variations in the hard power law and thermal 
components is negligible. 
 
We argue that the BH X-ray spectrum in the high-soft state is formed in the relatively 
cold accretion flow  with a subrelativistic bulk velocity $v_{bulk}\lax c$ and 
with  a temperature of a few keV and less ($v_{th}\ll c$).  In such a flow 
the effect of the bulk Comptonization 
is much stronger than
the effect of the thermal Comptonization which is a second order with respect to 
$v_{th}/c$.
In this {\it Letter} we present results of Monte Carlo simulations probing the  spatial, spectral, timing and time-lag  properties of X-ray radiation in   
CI atmosphere  in \S\S 2-4. Comparisons are drawn to the recent X-ray observations of BH sources.  
Summary and conclusions follow in \S 5. 
\section{Emergent spectrum and hard photon spatial distribution}  
The geometry used in these simulations is similar to the one used 
in the Laurent \& Titarchuk (1999), hereafter LT99, consisting of a
thin disk  with an inner radius of 3  $r_{\rm S}$, merged with  a spherical CI cloud   
harboring a BH in its center, where $r_{\rm S}=2GM/c^2$ is 
the Sc{h}warzchild radius and  $M$ is a BH mass.  The cloud outer radius is $r_{out}$.   
The disk is assumed always  to be optically thick.    
 
In addition to free-fall into the central BH, we have also taken   
into account the thermal motion of the  CI electrons,   
simulated at an electron temperature of 5 keV. This  is likely to be a typical 
temperature of the CI in the high-soft state of galactic BHs 
[see  Chakrabarti \& Titarchuk (1995) and BOR99 for details].
The seed X-ray photons were generated uniformly and isotropically    
 at the surface of the border of the accretion disk, 
 from $r_{d,in}= {3} r_{\rm S}$ to $r_{d, out} = 10r_s$. 
 These photons were generated according to a thermal spectrum with a
single temperature of 0.9 keV, similar to the   ones measured in Black Hole binary systems 
(see for example BOR99). In fact, BOR99 also demonstrate that the multicolor  disk spectrum 
can be fitted by the effective single temperature blackbody spectrum in the energy range of interest   
(for photon energies higher than 2 keV).
The parameters of our simulations are the BH mass $m$ in  solar units, 
the CI electron temperature, $T_e=5$keV,  
the mass accretion rate,  $\dot m=4$ in Eddington units (see LT99), 
and the cloud outer radius, $r_{out}=10r_{\rm S}$.  
It is worth noting that, for these  parameters, 
the bulk motion effects are not significant   at $r> 10r_{\rm S}$.\footnote{  
In general, the bulk motion Comptonization is effective when $\tau v_{bulk}/c
=\dot m r_{\rm S}/r$  {\it is not smaller than unity} (e.g. Rees, 1978; Titarchuk et al. 1997).}  

In Figure 1 we present the simulated spectrum of the X-ray emission emerging
from the CI atmosphere.  The spectrum exhibits  three features: a soft X-ray bump, an extended power law 
and a sharp exponential turnover near 300 keV. 
As demonstrated previously (Titarchuk \& Zannias 1998, hereafter TZ98; LT99) 
the extended power law is a result of soft photon  upscattering off CI electrons.
 The qualitative explanation of this phenomena was given 
by Ebisawa et al. (1996), hereafter ETC and later by Papathanassiou \& Psaltis (2001). 
The exponential turnover is formed by  a small fraction of those photons which undergo scatterings  
near the horizon where  the strong curvature of the photon trajectories  prevent us from  detecting most 
them.   In Figure 1 we indicate, with arrows  the places in the CI atmosphere where  photons 
of a particular energy mainly  come from.\footnote{Recently, Reig et al. (2001) show 
that the extended power law can be also formed as a result of the bulk motion Comptonization
when the rotational component is dominant in the flow. The photons gain energy from the rotational motion 
of the electrons.}
 
A {precise} analysis of the X-ray photon distribution in CI atmosphere can be made through  
calculation of the source function,  either by semi-analytical methods solving  
the relativistic kinetic equation (RKE) (TZ98) or by Monte Carlo simulations (LT99).  
TZ98 calculate the photon kinetics in the lab frame  demonstrating  that  the source function  has a strong peak near the photon bending 
radius, $1.5r_s$ (Fig. 3 there). 
In Figure 2 we show the Monte Carlo simulated source functions  for four energy bands:  
2-5 keV (curve a), 5-13 keV (curve b), 19-29 keV (curve c) and 60-150 keV (curve d). 
The peak at around $(1.5-2)r_{\rm S}$ is clearly seen 
for  the second and the third bands which is in a good agreement with TZ98's source function. 
\footnote{TZ98 implemented the method of separation of variables to solve the RKE Green's function. 
They then showed that the source function of the upscattered photons is distributed according to  
the first eigenfunction of the RKE space operator  defined by equations (21-24) in TZ98.} 
 
TZ98 calculated  the upscattering part of spectrum neglecting the recoil effect. 
Our Monte Carlo simulations, not limited by TZ98's approximation, reproduce    
the source function spatial distribution for high  energy bands (curve {\it d} in Figure 2).  
We confirm that the density  of the  highest energy X-ray photons is  concentrated  
near the black hole horizon. 
Comparison of TZ98's semi-analytical calculations and our Monte-Carlo simulations  leads us to conclude 
that the source function really follows the RKE  first eigenfunction distribution
 until  very  high energies. At that point  the upscattering photon spatial distribution also becomes  a function of  energy.    
As seen from Figure 2 
the source function in the soft energy band (curve {\it a}) has two maxima: 
one is at 2.2 $r_{\rm S}$ related to the photon 
bending radius and another (wide one) is at 5 $r_{\rm S}$ affected by the disk emission area. 
 
\section{An Illumination Effect of CI Atmosphere and High Frequency QPO Phenomena}  
 
We {remind} the reader that the emergent spectrum is a result of integration of the product of  
the photon escape probability and the source function distribution along a line of sight 
(e.g. Chandrasekhar 1960).  
The normalization of the upscattering spectrum is determined by the fraction of the soft (disk) 
 photons which  illuminate the inner region of the CI atmosphere below $3r_{\rm S}$, i.e.  
 around the maximum of the upscattered photon source function.     
From the Monte Carlo simulations we extract the fraction of the soft photons emitted  
 at some particular disk radius $r$  which  form the high energy part of spectrum 
 (for $E>10$ keV). This distribution is presented in the upper panel of Figure 3. 
As  seen from this plot, the strength of the high energy tail is mostly determined by the photons  
emitted at the inner edge of the disk. Thus any perturbation in the disk should be immediately translated 
to the oscillation of the hard tail with a frequency related to the inner disk edge.  
The CI atmosphere manifests the  high frequencies QPO of the innermost part of the disk.  
In the lower panel of Figure 3 we present the power density spectrum for the simulated X-ray emission coming 
from  CI atmosphere in the energies higher than 10 keV. We assume that the PDS is a sum of  
the red noise component  (where the power law index equals to 1) and QPOs power  proportional  
to the illumination factor  for a given disk radius, $r$ (see the upper panel). 
We also assume that the perturbation frequency  at $r$ is related to  the Keplerian frequency  
 $\nu_{\rm K}=2.2(3r_{\rm s}/r)^{1.5}/m$ kHz. 

There is a striking similarity between the high frequency QPO  and spectral energy distribution
 of the Monte Carlo results  
and real observations of BHC in their soft-high state. For example,  
in XTE J1550-564 the $\sim 200-$Hz QPO phenomenon tends to be detected in the high state at times 
when the bolometric luminosity surges and the hard-power-law spectral component is prominent 
(TS02). The noted lack of coherence between intensity variations of the      
high-soft-state low energy bands is also in a good agreement with the our simulations where  
the high energy tail intensity {\it correlates} with the supply of the soft photons 
from the inner disk edge but {\it it does   not correlate} with the production of the disk photons at large.   
 
\section{Positive and Negative  Time Lags} 
Additional important information related to the X-ray spectral-energy 
distribution can be extracted from the time lags between different energy 
bands. The hard and soft lags have been observed for several sources (e.g. Reig et al. 2000, hereafter
R00: Tomsick \& Kaaret 2000 for GRS 1915+105 and Wijnands et al. 1999, Remillard et al. 2001 for 
XTE J1550-564).  
It is natural to expect the 
positive time lags in the case of the unsaturated thermal Comptonization. 
The primary soft photons gain energy 
in   process of scattering off hot electrons; thus the hard photons spend more 
time in the cloud  than the soft ones. 
Nobili at al. (2000)
first  suggested that in a corona with a temperature stratification (a hot core and a relatively
cold outer part) the thermal Comptonization can account for the positive and 
negative time lags as well as the observed colors (see R00).\footnote {This  Comptonization model with a hot core (for the similar models,
see also Skibo \& Dermer 1995 and Lehr et al. 2000) may reproduce 
the intrinsic spectral and timing properties of the converging inflow.}  
  
In  the bulk-motion Comptonization case, the  soft disk  
photons at first gain energy in  the deep layers of the converging inflow,  
and then in their subsequent path towards the observer lose 
energy in the relatively cold outer layers. If the overall optical 
depth of the converging inflow atmosphere (or the mass accretion rate) is 
near unity, we would detect only the positive lags as in the thermal  
 Comptonization case, because relatively few photons would lose energy in  
 escaping. But with an increase of the 
optical depth, the soft lags appear because more hard photons lose their 
energy in the cold outer layers. 

In Figure 4  we present the  calculations of the 
 of time lags for two energy bands 2-13 keV and 13-900 keV. 
We first compute the time spent by the photon, drawn in a given direction,  
to cross the whole system without being scattered.  
Then, we compute the real time of flight  of the photon, taking into account  
its scattering and its complete trajectory in the system. The general relativistic  
gravitational time dilatation was also taken into account  in these computations. 
The time lag is then  determined as a difference between these two times. 
In this time-lag definition, a photon which undergoes no scattering has a zero lag.  
Thus the time lags between two energy bands can be defined a difference of these 
two time lags, i.e. $\delta t =t_2-t_1$ where $t_1$ and $t_2$ are time lags  
for 2-13-keV and 13-900-keV bands respectively. 
The time lags for 2-13-keV band are distributed over the interval  between 1 and 600  
Schwarzschild times,  $t_{\rm S}=r_{\rm S}/c$ and  the time lags for 13-900-keV  
band are distributed over the interval  between 200 and $10^3$ ${t}_{\rm S}$. 
For a ten solar mass BH  $t_{\rm S}=0.1$ ms.  
Thus, for $t_1<15$ ms all time lags $\delta t$ are positive, i.e. high energy  photons  
are produced later than the soft energies photons.  The $t_2-$ distribution has a broad  
 peak at 30 ms and $t_1-$ distribution has a broad peak at 8 ms.     
Consequently,  the positive lags  $\delta t$ are mostly  higher than 20 ms. 
which is in a good agreement with the time lags detection in GRS 1915+105 (R00).  
The  values of the positive time lags for this source are consistent with the BH 
mass being around 20 solar masses (see BOR99 for the BH mass  determination in GRS 1915+105).  
 We show that the absolute value of the  
time (phase) lags get  higher with the energy. In fact, Wijnands et al. (1999) 
for XTE J1550-564 (see also  Remillard et al. 2001)  and Tomsick \& Kaaret (2000) 
for GRS 1915+105 show  that the phase lags of both signs   
increase from 0 to 0.5 radians as the photon energy varies from 3 keV to 100 keV.     
These phase lag changes correspond to the time lag changes from 0 to 15 ms and from 0 to  
50 ms for XTE J1550-564 and GRS 1915+105 respectively.     
These values are very close to what we obtain in our simulations.  
 
For $t_1>15$ ms we get  the time lags of both signs. Positive values of $\delta t$  
should be of order  30 ms and higher. Negative values should not be higher than 20 ms  
for a 10 solar mass BH. 
 Our simulations demonstrate    that for high accretion  
rates  time lags of both signs are present, i.e. the upscattering and down-scattering are 
equally important in the formation of CI spectrum, whereas for the low mass accretion rates 
the upscattering process  leads to the CI spectral formation and only the positive times lags are  
present.
Our simulations are in a good agreement  with the R00's values of $\delta=30-60$ ms.  
The results for the simulated time lags can be understood using the simple 
upscattering model (see ETC). We assume that  
at any scattering event the relative mean energy change is constant, $\eta$, i.e. 
$\eta=<\delta E>/E$. The number of scatterings $k$ which are needed for the soft photon of energy  
$E_0$ to gain energy E is  $k(E)=\ln(E/E_0)/\ln(1+\eta)$. 
Then $\delta t= k\times l(1-r_{\rm S}/r)^{-1}/c$ 
where $l$ is the photon mean free path, $l=[b(E)\sigma_{\rm T}n]^{-1}$, $(1-r_{\rm S}/r)$ is 
 the relativistic  dilatation factor (e.g. Landau \& Lifshitz 1971)  
and $b(E)$ is an energy dependent factor less than unity (e.g. Pozdnyakov et al, 1983).     
The free path $l$ is estimated as $l\sim 2(r/r_{\rm S})^{1.5}r_{\rm S}/b(E)\dot m$ with an assumption that
the number density $n$ is calculated for the free fall distribution, (see LT99, Eq. 2).  
With $r/r_{\rm S}\approx 1.18$ (see Fig. 2, curve {\it d}), $\dot m=4$ and  $b\approx0.5$ for $E=100$ keV  
we get $l\approx 1.2 r_{\rm S}$.  
$b(E)\approx 1$ for $E=10$ keV   and $r/r_{\rm S}\gtorder 2$ and consequently
$l\gtorder 1.5 r_{\rm S}$.\footnote{It is worth noting that the emergent high energy photon  gains its energy being  
predominantly close to horizon, $r=(1-2)r_{\rm S}$ where the photon trajectory are close to  
circular (e.g. TZ98). Thus our estimated values of the free path  $l\sim (1-1.5) r_{\rm S}$ are consistent 
with the trajectory length. With $\eta=0.1$ extracted from our  simulations for $E<100$  keV 
where the extended power law is seen in the spectrum (see also Papathanassiou \& Psaltis 2001, Fig. 3)
the time lags  are $\delta t= 7.6$ ms and $50$ ms for $E=10$ keV and $100$ keV respectively.} 
     
\section{Summary and Conclusions} 
 
Using Monte Carlo simulations we have presented a detailed timing analysis  
of  the X-ray radiation from a CI atmosphere.  
 (1) We confirm TZ98's results that  the high energy photons are produced predominantly in  
the deep layers of the CI atmosphere from 1 to 2 Schwarzchild radii. 
(2) We also confirm that the $~$ 100-Hz phenomena should be seen in the high energy tail of the spectrum 
rather than in the soft spectral component because the inner part of the converging inflow  
is mostly fed by soft 
photons from the innermost part of the disk but the contribution of this disk area to the disk emission is 
small.  
(3)  We find that the characteristic time lags  in the converging inflow are order of  a few tens of ms  
at 10 keV and the hard (positive) time lags increase logarithmically with energy.  
(4) We demonstrate that soft (negative) along with  hard (positive) lags are present 
in the X-ray emission from the converging inflow for the mass accretion higher than 
Eddington.  
In the end we can also conclude that that almost all of the X-ray  timing and spectral properties 
of  the high-soft state in BHs tends to indicate that {\it the high energy tail of spectrum 
is produced within the CI region, $(1-3)\times r_{\rm S}$  
where the bulk inflow (gravitational attraction) is a dominant process}.      
 
We acknowledge the fruitful discussions with Chris Shrader and Paul Ray. 
We also acknowledge the useful suggestions and the valuable corrections by the referee 
which improve the paper presentation.

\clearpage

\begin{figure} 
\epsscale{0.7} 
\plotone{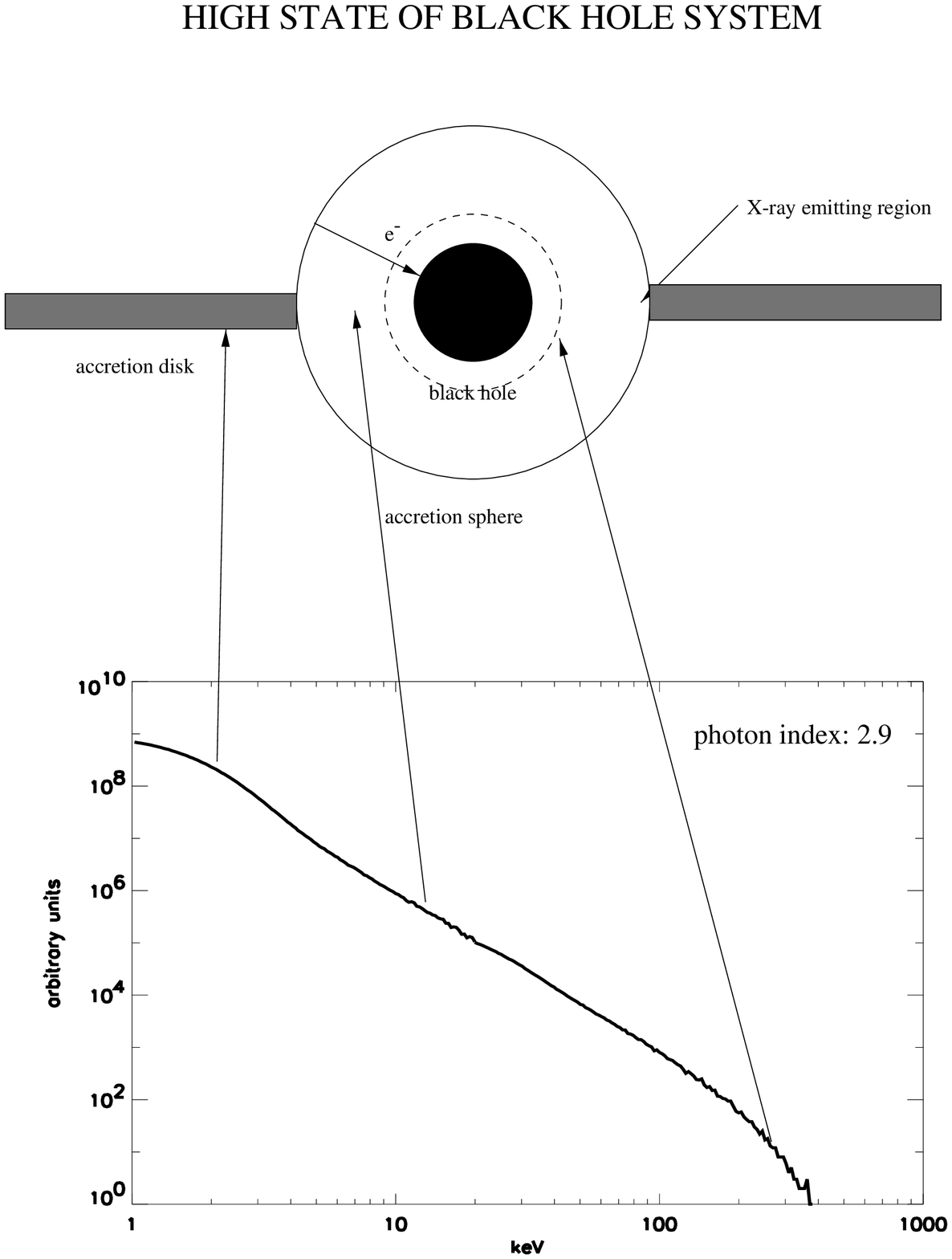} 
\caption{Monte Carlo simulated emergent spectrum of Converging Inflow.  
Mass accretion rate in Eddington units, $\dot m=4$, electron temperature 
of the flow $kT_e={5}$ keV. } 
\label{fig1} 
\end{figure} 
 
\begin{figure} 
\epsscale{0.7} 
\plotone{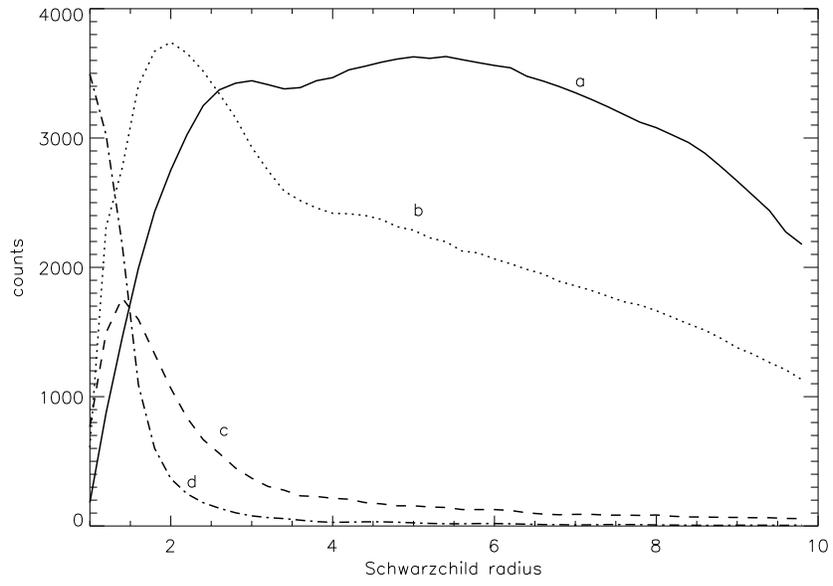} 
\caption{ Source photon spatial distribution in the converging inflow atmosphere for different energy bands: 
{\it a} is  for (2-5) keV, {\it b} is for (5-13) keV, {\it c} is for (19-29) keV and 
{\it d} is for (60-150) keV. } 
\label{fig2} 
\end{figure} 
 
\begin{figure} 
\epsscale{0.7} 
\plotone{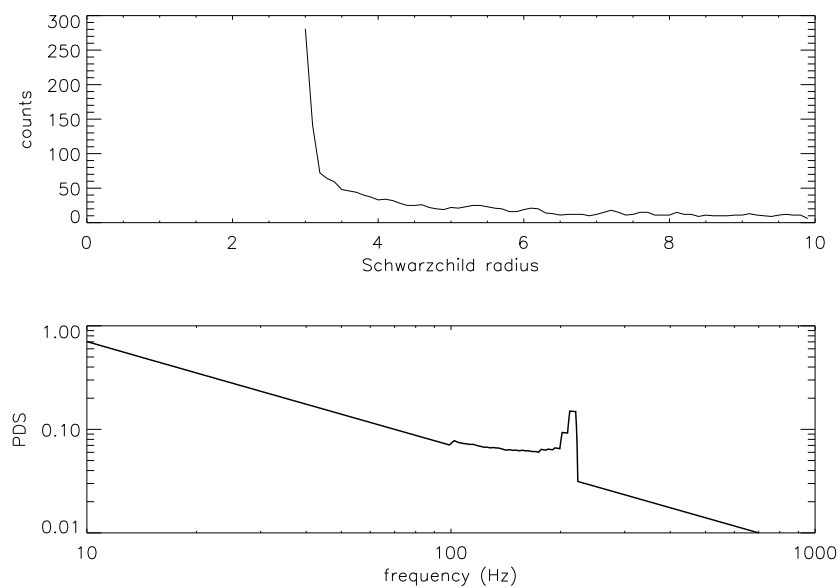} 
\caption{ Distribution of soft photons over disk radius  which upscatter  to energies 10 keV and higher  
in  CI atmosphere (upper panel). Power density spectrum for photon energies higher than 10 keV  
(lower panel). It is assumed that any disk annulus oscillates with Keplerian frequency. The continuum 
is a red noise.} 
\label{fig3} 
\end{figure}

\begin{figure} 
\epsscale{0.7} 
\plotone{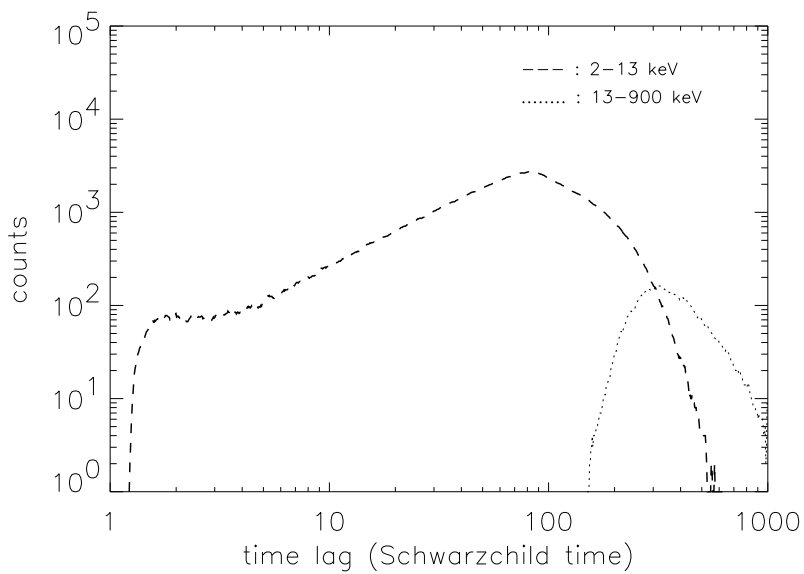} 
\caption{ Time lag distributions for two {energy bands}. One can  clearly see the area 
between 150 and 600 $r_{\rm S}/c $ where time lags of both signs can be present. } 
\label{fig4} 
\end{figure} 
 
\clearpage 
 
\end{document}